\documentstyle[preprint,aps,floats,epsf]{revtex}

\def\mw{m_W}
\def\mz{m_Z}
\def\ms{m_{\tilde{q}}}

\def \7#1#2 {\mathop{\null#2}\limits^{#1}}   
\def \5#1#2 {\mathop{\null#2}\limits_{#1}}   

\begin{document}


\preprint{
\font\fortssbx=cmssbx10 scaled \magstep2
\hbox to \hsize{
\hfill$\vcenter{\hbox{\bf IUHET-313}
                \hbox{\bf MADPH-98-1064}
                \hbox{\bf hep-ph/9809240}
                \hbox{September 1998}}$ }
}

\title{\vspace*{1cm}
Production of $Z$ Boson Pairs via Gluon Fusion \\
in the Minimal Supersymmetric Model}

\author{M.S.~Berger$^a$ and Chung Kao$^b$}

\address{
$^a$Department of Physics, Indiana University, Bloomington, IN 47405, USA \\
$^b$Department of Physics, University of Wisconsin, Madison, WI 53706, USA }

\maketitle

\thispagestyle{empty}

\begin{abstract}

We present the full one-loop calculation for $gg \to ZZ$ 
in the Minimal Supersymmetric Model (MSSM) 
including nonresonant contributions from the squark loop diagrams 
and provide analytical expressions for the helicity amplitudes. 
The one-loop process $gg \to ZZ$ via quark loops 
has been calculated in the Standard Model. 
In supersymmetric models, additional contributions arise from squark loops. 
In some regions of the MSSM parameter space, 
the top and bottom squark loops can make important contributions 
to the diagrams involving Higgs bosons. 
The heavy Higgs scalar ($H$) might be detected at the Large Hadron Collider 
via $gg \to H \to ZZ$ for $\tan\beta \alt 5$.

\end{abstract}

%
 

\section{ Introduction }

One of the main purposes for constructing a hadron supercollider 
is to discover the mechanism responsible for the spontaneous breaking 
of the electroweak symmetry. 
A large amount of attention has been devoted recently to the ability 
of detecting a Higgs boson at the CERN Large Hadron Collider (LHC). 
Gluon fusion is the main mechanism of producing Higgs bosons at the LHC, 
and if a CP-even Higgs scalar is sufficiently heavy it can decay 
into $Z$ boson pairs. 
In the Standard Model, this is a useful detection channel 
for the Higgs boson with a mass $M_{H_{SM}} \agt 2M_Z$ since the $H_{SM}ZZ$ 
coupling grows with Higgs boson mass.

The physical Higgs bosons of the minimal supersymmetric extension
to the Standard Model (MSSM) is comprised of two CP-even states, 
a lighter $h$ and a heavier $H$, one CP-odd state, $A$, and two charged 
Higgs bosons $H^\pm$. The Higgs potential is constrained by supersymmetry 
such that all tree-level Higgs boson masses and couplings are determined 
by just two independent parameters, commonly chosen to be the mass 
of the CP-odd pseudoscalar ($m_A$) and the ratio of vacuum expectation 
values (VEVs) of Higgs fields ($\tan\beta \equiv v_2/v_1$).
The CP-even states can couple to the $ZZ$ final state
at the tree level, but only the heavier one ($H$) can be massive enough 
to decay into on-shell $Z$ bosons. 
For $m_H \sim m_A \agt 2m_t$, discovery of the MSSM heavy Higgs scalar 
in the $ZZ$ channel could be problematic due to the fact 
that the Higgs couplings to the gauge bosons arise from the D-term 
contributions to the scalar potential. As the $H$ becomes heavy, 
its coupling with the gauge bosons becomes smaller.
Previous studies have found that the production rate for $gg\to H \to ZZ$ 
is large enough to make this channel a viable one for Higgs detection 
provided $\tan\beta \equiv v_2/v_1$ is relatively small ($\alt 5$) and 
that the Higgs boson is not too heavy ($m_H \alt 350$~GeV) 
\cite{bbps,Z2Z2,LHC,gsw}.
The impact of supersymmetric decay modes has also been 
investigated \cite{Z2Z2}.

The production cross section of $gg \to \phi$ ($\phi = H, h, A$, and $H_{SM}$) 
has been studied in increasing detail in the last few years, 
with much attention devoted to the Standard Model Higgs boson 
or to the lighter MSSM Higgs scalar, $h$. 
The impact of squark loops in the production process has been
included, and the impact of squark mixing has been investigated \cite{Kileng}.
The QCD corrections to the production process have been calculated and shown
to give a large enhancement to the cross section \cite{Spira}.

It is well known \cite{ggzz1,ggzz2} that the dominant background to the channel
$gg\to H_{SM}\to ZZ$ in the Standard Model is $q\overline{q} \to ZZ$, and the 
one-loop continuum production of $Z$ pairs via quark loops, $gg\to ZZ$.
In the MSSM the existence of the scalar partners to the quarks gives 
a new contribution to the continuum production that can interfere 
with the quark loops and possibly affect the overall level of background. 
If the masses of the squarks are much larger than the electroweak scale, 
their contributions to the signal should be small as their mass arises 
from soft terms rather than from the Higgs mechanism. 
For squarks with masses comparable to the electroweak scale, 
significant contributions to the Higgs production signal might be expected.
It has been shown \cite{Kileng,Konig,Djouadi} 
that the squark loops can have a significant impact 
on the production cross section if the squarks are fairly light. 
Therefore, a natural step is to also investigate the impact of the 
squark loops 
on the background.

The Feynman diagrams for the background coming from squark loops 
to the Higgs signal in the channel $gg \to H \to ZZ$ is shown in Fig.~1. 
Since only the third generation gives a sizable contribution to the 
resonant Higgs contribution $gg\to H\to ZZ$, one might expect an important
contribution from the continuum background for which all three generation
squarks are expected to contribute. 
The central issue concerning the detectability of the Higgs boson
is whether the squark loop contribution can be comparable 
to the quark loop contribution for energies near the Higgs mass. 
The squark loop contribution is maximal near the production threshold
($\sqrt{s}=2m_{\tilde{q}}$), so we are most interested in the size of 
the squark loop background 
for $2m_{\tilde{q}}\approx M_H$.

Another potential application of the process under consideration is a 
measurement of the $Hgg$ coupling. 
The production mechanism involves a coupling to the Higgs boson (a Yukawa 
coupling) that implies that only the third generation matter make a sizeable
contribution to the rate. Each generation contributes equally to the 
background (assuming equal masses for squarks in different generations), so 
potentially there is an enhancement as there is in the case of quark loops. 
A large background from squark loops could conceivably alter the regions of 
parameter space (the $M_A-\tan \beta$ plane) for which the heavy Higgs can be 
discovered. 

The gluon coupling to the Higgs boson is mediated by triangle graphs. 
The calculation  of the background involves box graphs and is significantly 
more complicated than that of the Higgs production cross section.
The one-loop integrals, which are expressible 
in terms of Spence functions (dilogarithms) \cite{Tini,FF}, 
were evaluated numerically using a FORTRAN code \cite{LOOP}.
We have not included squark mixing in this first survey, 
and leave the more general case to a future paper \cite{prep}. 
However we do not expect squark mixing to have a significant effect 
on the cross sections. While the cross section for the resonant Higgs
graphs can be significantly increased by an enhanced $H\tilde{t}_1\tilde{t}_1$
coupling \cite{Djouadi}, squark mixing 
results only in mixing angle factors in the couplings of squarks to the $Z$ 
bosons and splits the degeneracy of the squark masses for the 
nonresonant graphs. 
One expects an enhanced contribution only from the loops 
involving $\tilde{t}_1$ squarks which might have a suppressed mass 
with significant mixing. 
If the lightest squark is less than half the mass 
of the Higgs boson, then the Higgs boson can decay into squark pairs 
making the Higgs broader and more difficult to detect in the $ZZ$ final state. 

\begin{center}
\epsfxsize=1.7in
\hspace*{0in}
\epsffile{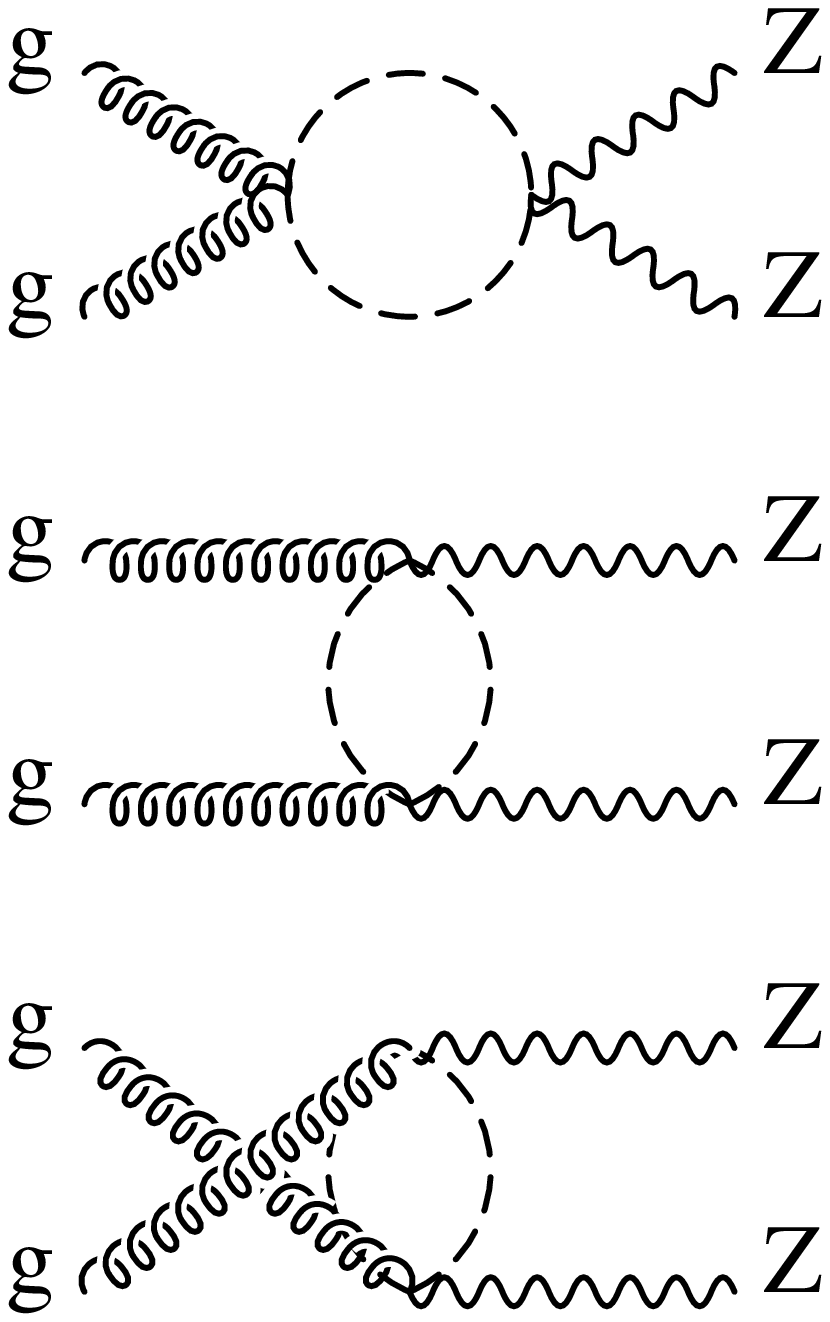}

\smallskip
\parbox{5.5in}{\small  Fig.1(a). The bubble diagrams.}
\end{center}

\begin{center}
\epsfxsize=1.7in
\hspace*{0in}
\epsffile{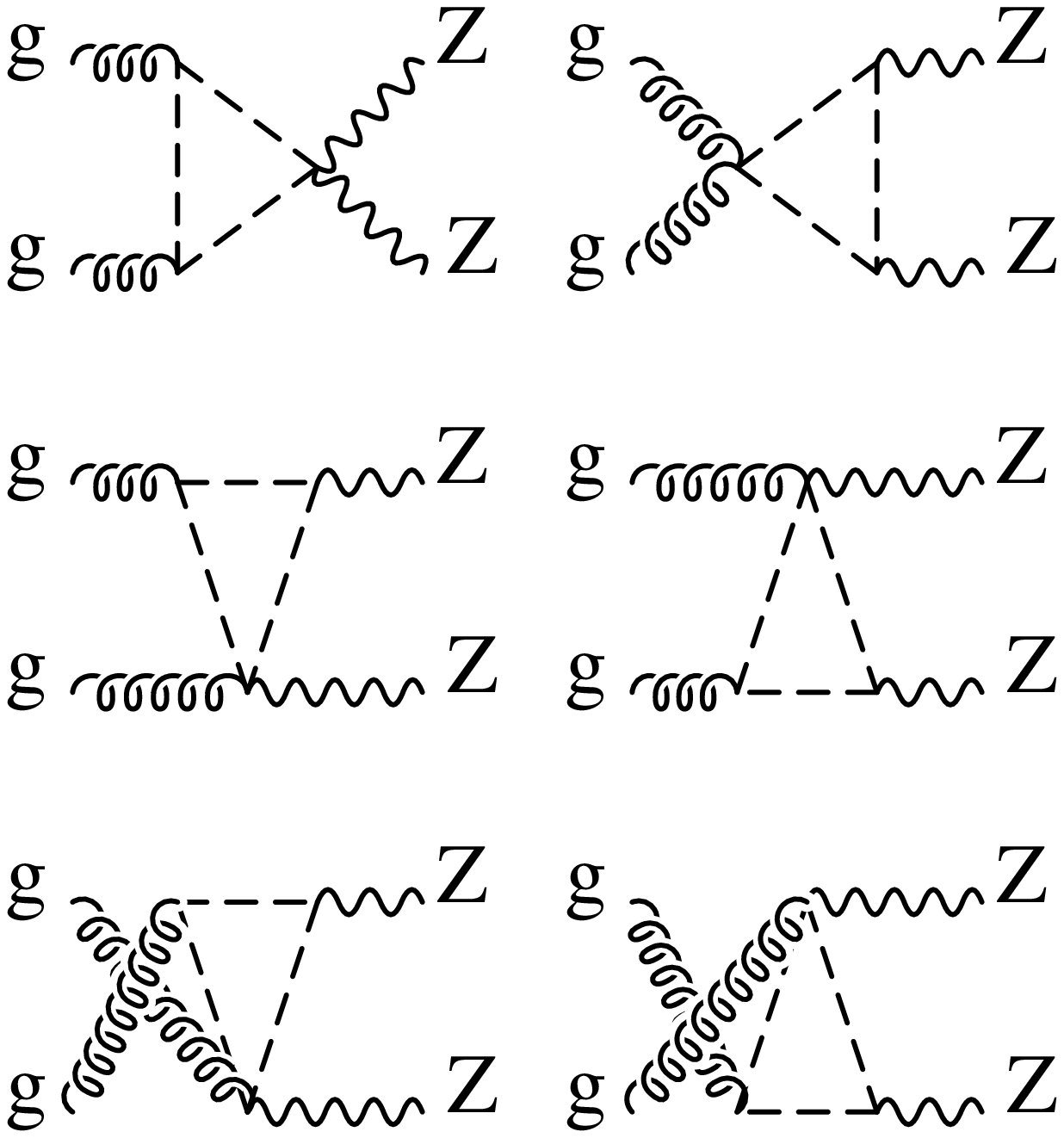}

\smallskip
\parbox{5.5in}{\small  Fig.1(b). The triangle diagrams.}
\end{center}

\begin{center}
\epsfxsize=1.7in
\hspace*{0in}
\epsffile{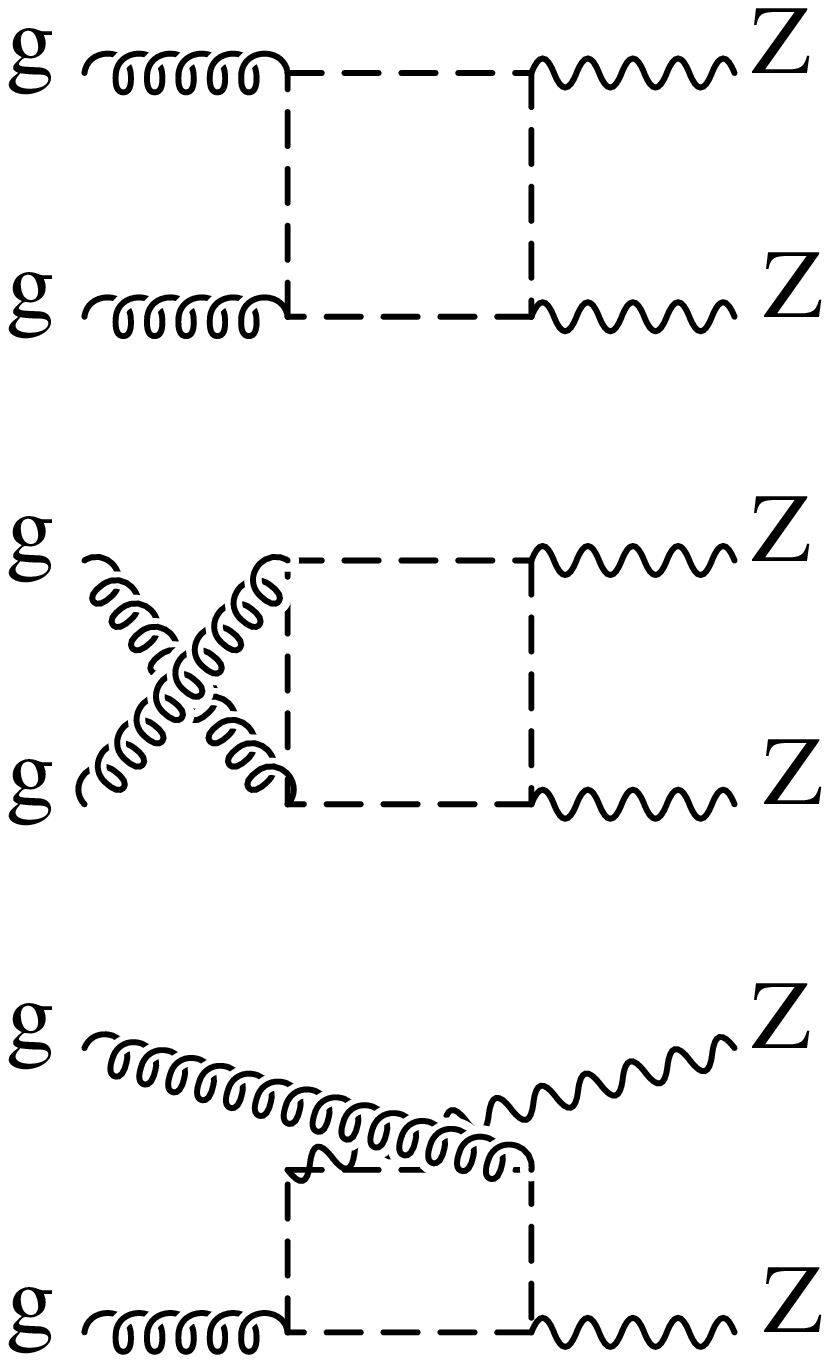}

\smallskip
\parbox{5.5in}{\small  Fig.1(c). The box diagrams.}
\end{center}

\begin{center}
\epsfxsize=2.2in
\hspace*{0in}
\epsffile{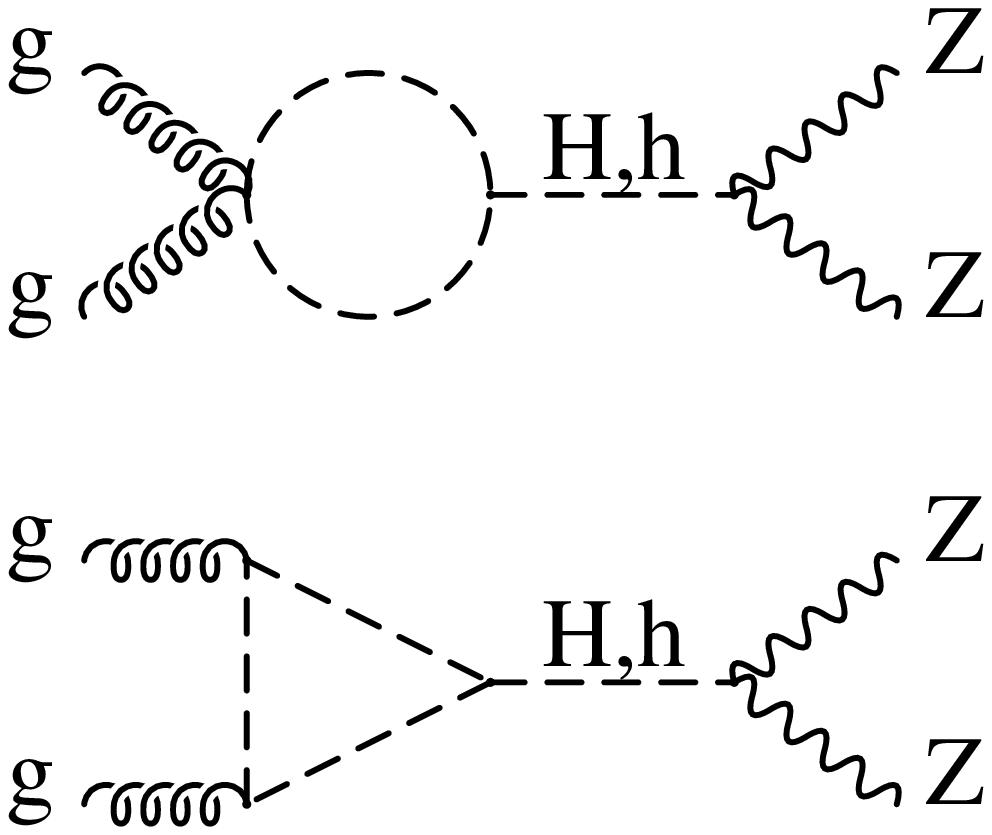}

\smallskip
\parbox{5.5in}{\small  Fig.1(d). The diagrams involving the Higgs bosons.}
\end{center}

\section{ The squark loop contribution to the background }

In this section we compare the size of the squark loop background
with that arising from the quark loops.
In the Standard Model, 
only quark loops contribute to the production of $ZZ$ 
from gluon fusion $gg \to ZZ$. 
In the MSSM there are additional contributions from squark loops, 
which typically are thought to be suppressed by large squark mass. 
In this section we present the cross sections expected at the LHC 
with $\sqrt{s}_{pp}=14$~TeV.  
In our numerical computations we use 
the CTEQ3 parton distribution functions \cite{CTEQ} 
with $\Lambda^4 = 0.177$ GeV. 
We take $\alpha ^{-1}=128$, $\sin ^2\theta _W=0.2319$, 
$\alpha _s(M_Z)=0.124$ and $m_t=175$~GeV.

In Figure 2 we plot the invariant mass distribution  
$d\sigma/dM_{ZZ}(pp \to ZZ +X)$ from $gg \to ZZ$ via the squark loops 
in the MSSM, assuming their masses are all degenerate 
at values of 100~GeV, 300~GeV and 500~GeV. 
The lightest squark considered is excluded by the Tevatron data \cite{squark}, 
but we include it for comparison purposes.
There is a rapid rise in the cross section until the threshold of real 
squark pair production after which the cross section falls again. 
Since the squark loop cross section peaks at $\sqrt{s}\sim 2m_{\tilde{q}}$, 
the largest contribution to the background from the squarks comes 
at subprocess ($gg\to ZZ$) energies slightly 
below the squark production threshold. 
The contribution to $gg\to ZZ$ from graphs with squark loops 
not involving the Higgs boson is separately gauge invariant, 
and the cross section falls off rapidly as $\sqrt{s} $ increases.

In Figure 3 we show the squark loop cross section for the polarization states
$LL$, $TT$ and $TL+LT$ of the $ZZ$ final state. The contribution 
from squark loops is always smaller than that from quark loops 
even for squarks as light as 100~GeV. The transversely polarized case 
dominates over the other modes for the squark loops just as it does 
for the contributions from the quark loops.

\begin{center}
\epsfxsize=4.5in
\hspace*{0in}
\epsffile{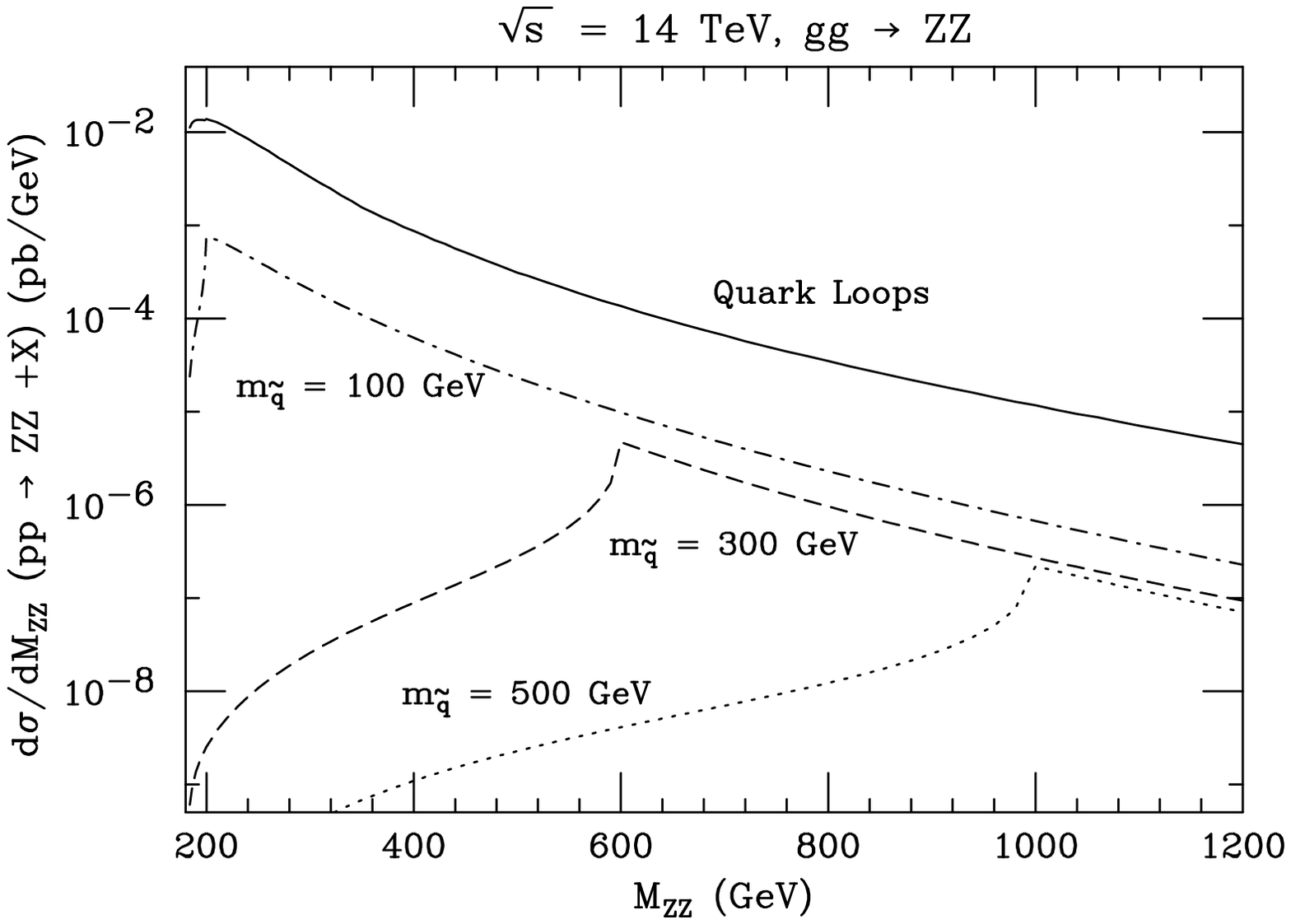}
\vspace*{-3.0in}

\smallskip
\parbox{5.5in}{\small  Fig.~2. $d\sigma /dM_{ZZ} ( pp\to ZZ+X )$, 
from $gg \to ZZ$, 
via squark loops with 12 squarks,
without Higgs bosons,
for $m_{\tilde{q}} = 100$ GeV, 300 GeV and 500 GeV.
Also shown are the contributions from the box diagrams 
of quark loops. These contributions are independent of 
$\tan \beta$ for fixed squark mass $m_{\tilde q}$.}
\end{center}

\begin{center}
\epsfxsize=4.5in
\hspace*{0in}
\epsffile{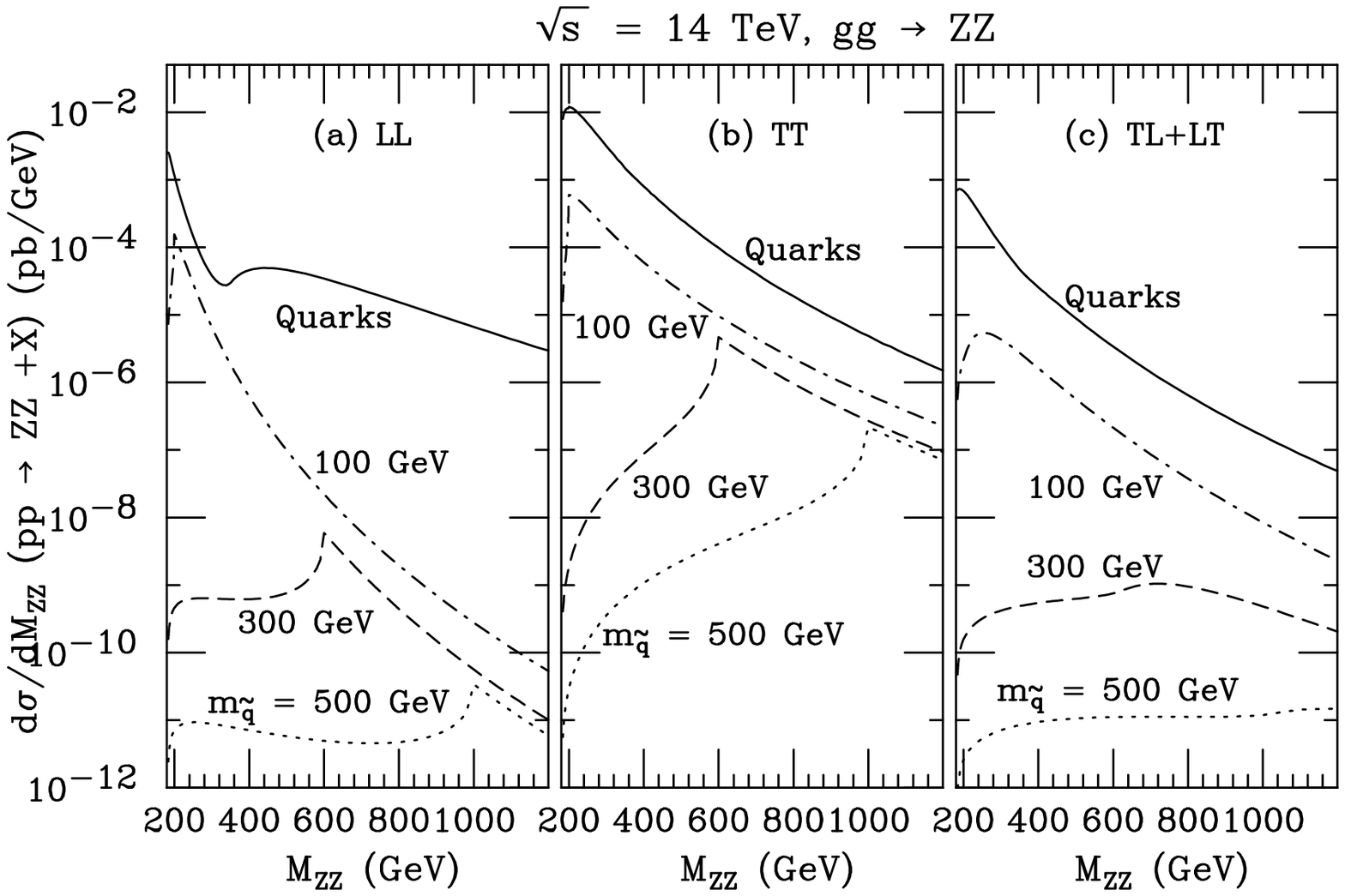}
\vspace*{-3.2in}

\smallskip
\parbox{5.5in}{\small  Fig.~3. $d\sigma /dM_{ZZ} ( pp\to ZZ+X )$, 
from $gg \to ZZ$, via squark loops with 12 squarks,
without Higgs bosons,
in the (a) $LL$, (b) $TT$ and (c) $TL+LT$ states of the $ZZ$
for $m_{\tilde{q}} = 100$ GeV, 300 GeV and 500 GeV.}
\end{center}

\section{ The supersymmetric limit }

Some contributions to the one-loop amplitudes for squark loops given in the
Appendix are closely related to parts of the helicity amplitudes for the 
quark loops.
These types of supersymmetric relationships \cite{susy1} 
were observed in the explicitly computed weak interaction process 
$Z\rightarrow 3 \gamma$ 
or equivalently $\gamma \gamma \to \gamma Z$ \cite{susy2}, and for 
$\gamma \gamma \to \gamma \gamma$ \cite{Jikia}. One can also compare
the contributions from fermion loops in the process 
$\gamma \gamma \to HH$\cite{ppHH} to the contribution that was obtained 
later for the gauge boson loops \cite{HH}. However the most numerically 
significant contributions to the cross section is often those contributions
that are not given by the supersymmetric relationship. In fact the asymptotic
behavior ($\sqrt{s}\to \infty$) is governed by the spin of the particles
exchanged in the $t$- and $u$-channels. For example, one has the dominance of
the $W$ boson loops over fermion 
loops in the process $\gamma \gamma \to ZZ$ \cite{ppzz}.

\begin{center}
\epsfxsize=4.5in
\hspace*{0in}
\epsffile{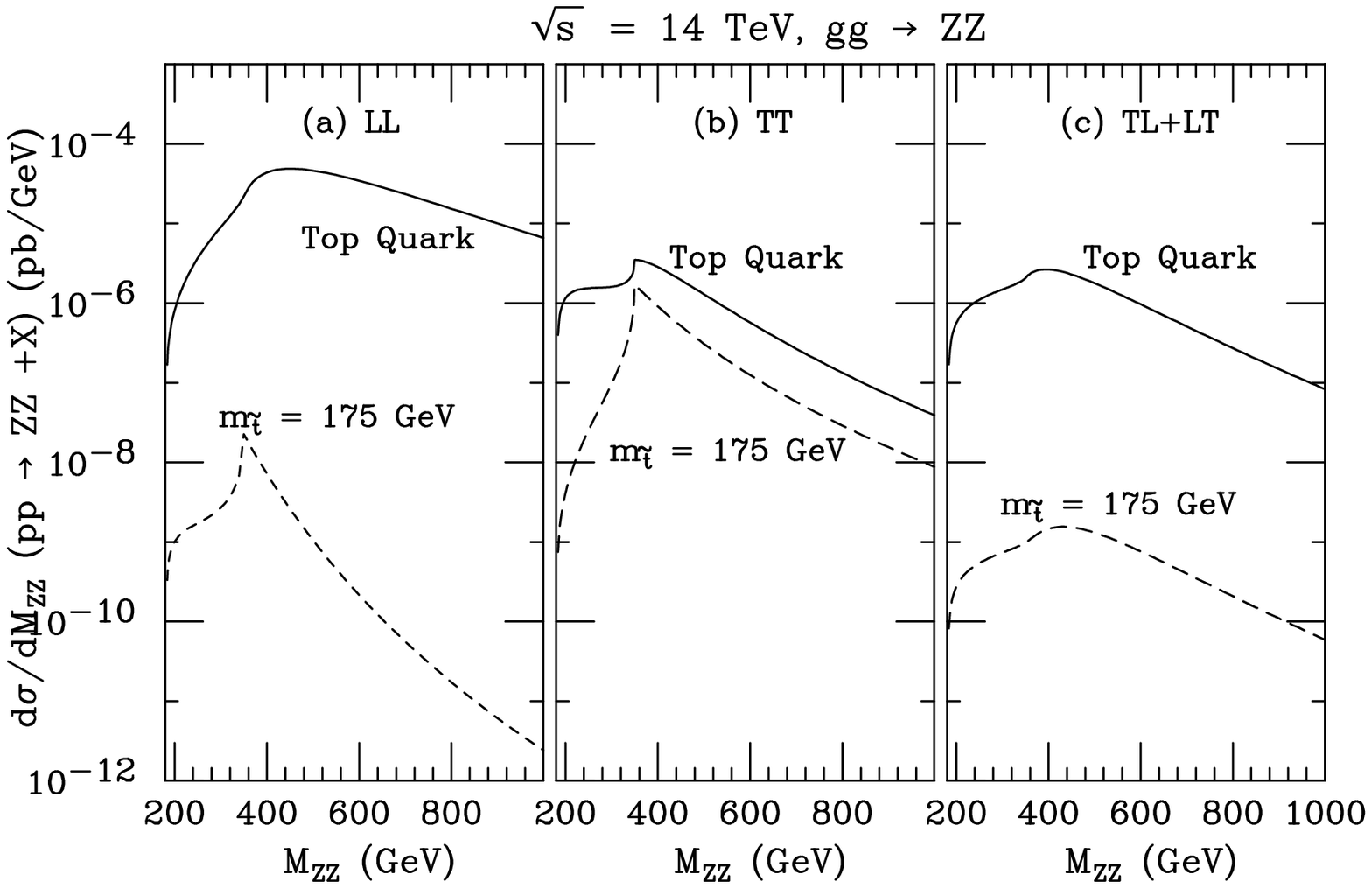}
\vspace*{-3.0in}

\smallskip
\parbox{5.5in}{\small  Fig.~4. Differential cross section of $pp \to ZZ +X$ from $gg \to ZZ$ 
via top quark and top squark loops without Higgs bosons 
for $m_t=m_{\tilde{t}}=175$~GeV.}
\end{center}
\vspace*{0.2in}

For the process $gg\to ZZ$ discussed in this paper, the SUSY relationship 
exists between the squark loop amplitudes and the parts of the quark loop 
amplitudes arising from the vector coupling of the $Z$ boson to quarks.
This relationship provides another consistency check on our analytic results
in addition to the usual check of gauge invariance.
In Fig.~4 the contributions from the top quark loops is compared to the 
contributions from $\tilde{t}_L$ and $\tilde{t}_R$ loops for masses set
equal to $m_t = m_{\tilde{t}} = 175$ GeV.

\section{ Interference effects }

In Fig.~5 we present the contributions to $gg\to ZZ$ including 
both the resonant Higgs graphs and the nonresonant graphs 
for the quark loops alone and for the squark loops alone 
with $\tan \beta=2$ and 10. 
The squark mass is 200~GeV, and $M_A$ is 300 GeV for
which $M_H \simeq 309$ GeV. For $\tan \beta > 5$ the differential cross 
section appears the same apart from the size of the Higgs boson resonance.

\begin{center}
\epsfxsize=4.5in
\hspace*{0in}
\epsffile{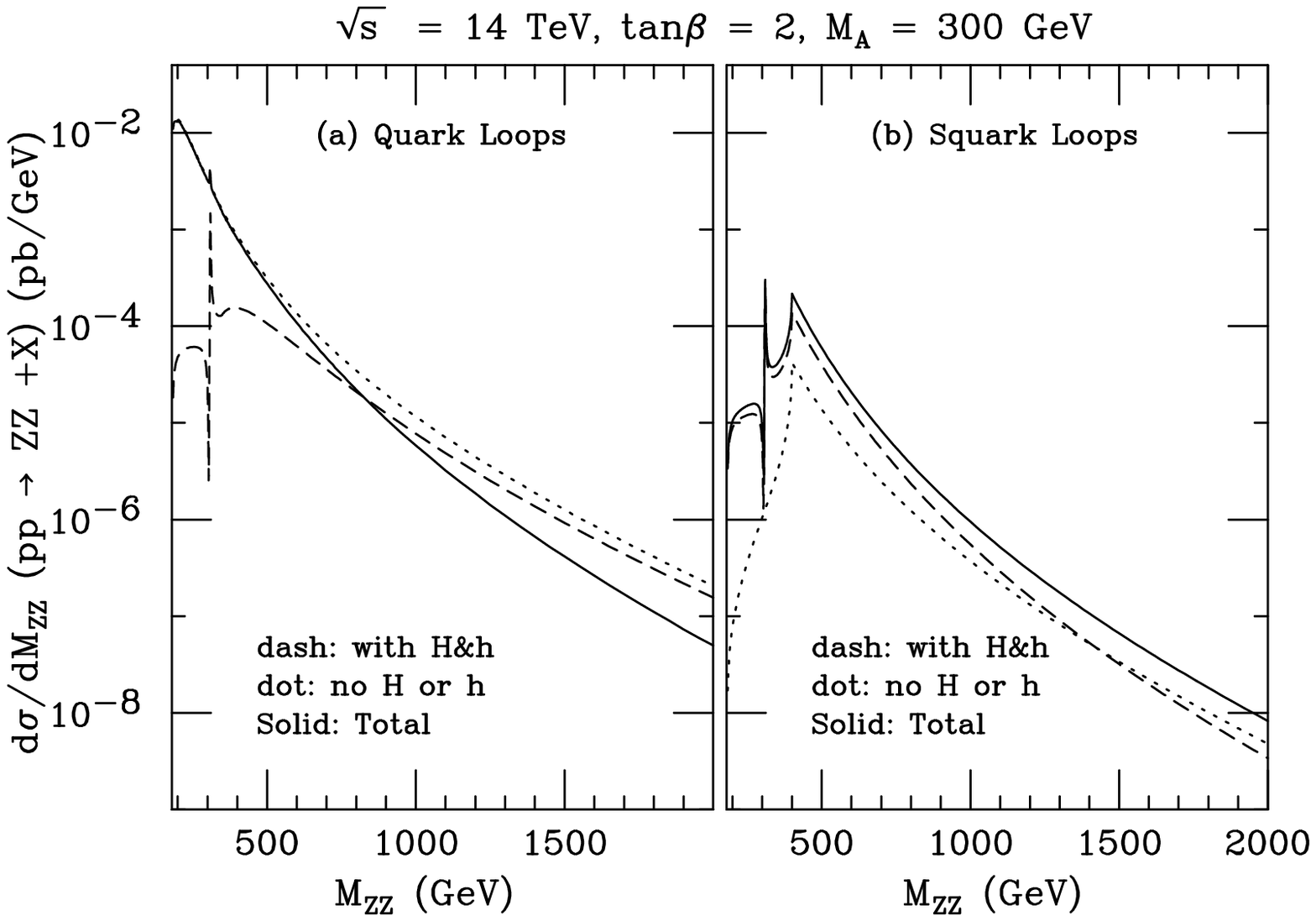}
\vspace*{-2.6in}

\smallskip
\parbox{5.5in}{\small  Fig.~5. Differential cross section of the Higgs signal 
in $pp$ collisions from (a) quark loops and (b) squark loops. The dashed lines
are the contributions from the Higgs graphs, the dotted lines are the 
contribution from the nonresonant graphs, and the solid lines are the total.}
\end{center}
\vspace*{0.2in}

An interesting property of the quark loop amplitudes 
is the cancellation of the Higgs graphs with the box diagrams 
which can be related to the good high energy behavior of 
$t\bar{t} \to ZZ$. 
The relative sign between the triangle and the box diagrams 
has been checked with the unitarity condition. 
We can cut the loop diagrams and obtain the tree processes 
$t\bar{t} \to ZZ$ and $b\bar{b} \to ZZ$.
Unitarity then requires a cancellation among tree diagrams at high energy.
No such cancellation is required for the scalar loops 
between diagrams with and without Higgs bosons.

\vspace*{-0.4in}
\begin{center}
\epsfxsize=5.2in
\hspace*{-0.3in}
\epsffile{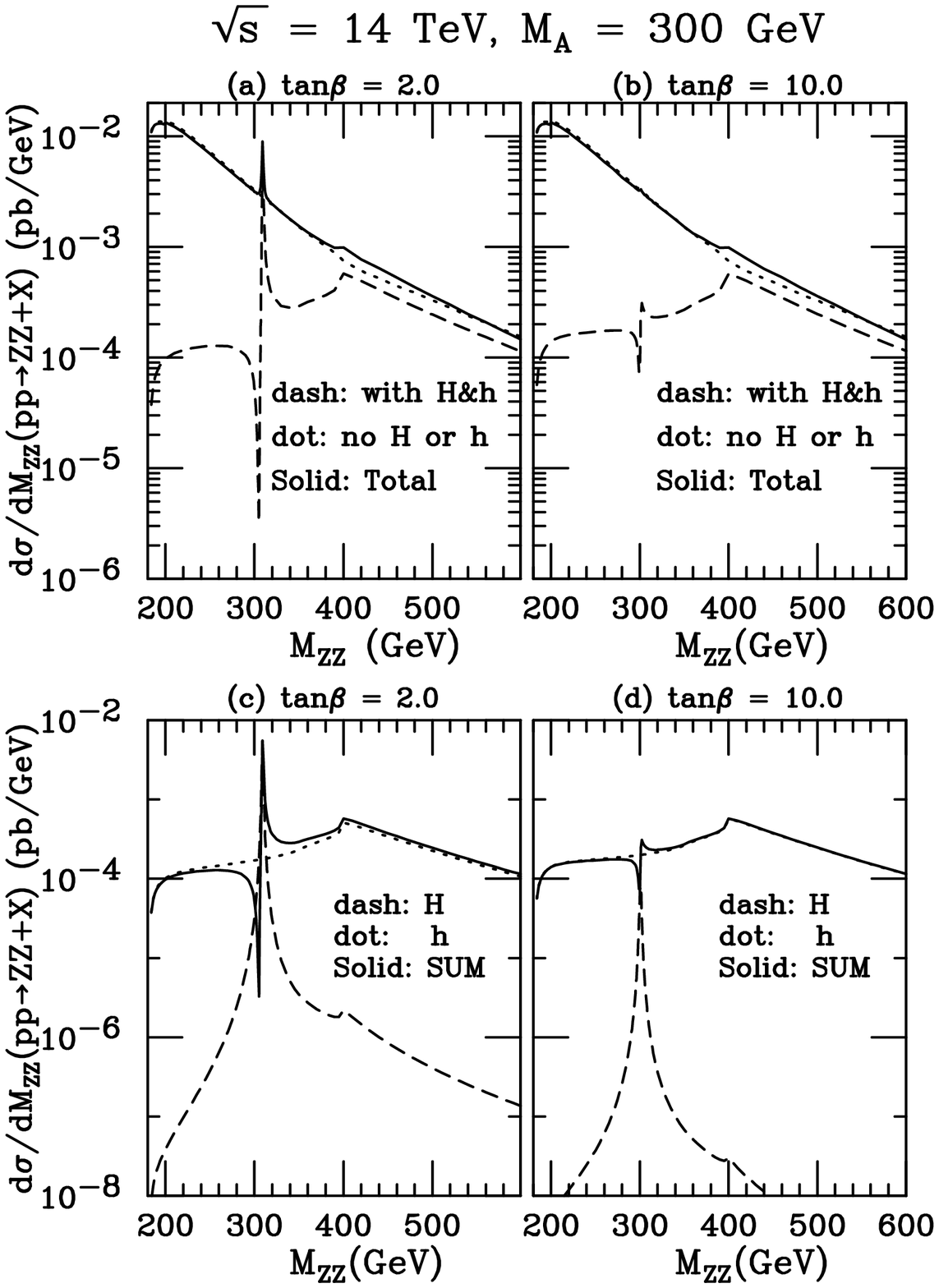}
\vspace*{-1.0in}

\smallskip
\parbox{5.5in}{\small  Fig.~6. Differential cross section of $pp \to ZZ +X$ 
from $gg \to ZZ$ 
via both quark and squark loops, for $m_{\tilde{q}} = 200$ GeV,
as a function of $M_{ZZ}$, 
for (a) $\tan\beta = 2$ and (b) $\tan\beta = 10$. 
Also presented is the destructive interference between diagrams 
with the $H$ and the $h$ 
for (c) $\tan\beta = 2$ and (d) $\tan\beta = 10$.}
\end{center}

In Fig.~6 we have plotted the contributions to $gg\to ZZ$ including 
both the resonant Higgs graphs and the nonresonant graphs 
for $\tan \beta=2$ and 10 including both the quark loops 
and the squark loops. The background from squark loops can be safely neglected 
around the Higgs pole region.

The $H$ resonance exhibits destructive interference just below the peak. 
This arises from interference with the contribution from the diagrams 
involving the light Higgs boson $h$. 
This feature is present for squark loops alone and also for quark loops alone, 
and should be present in nonsupersymmetric extensions of the 
Standard Model with extended Higgs sectors. 
In the MSSM it seems unlikely that this interference dip can be 
seen experimentally given the size of the underlying background.
We note that the contributions to $gg\to ZZ$ 
from squarks loops without Higgs bosons 
are also valid in nonminimal supersymmetric models since they do not
involve contributions from the Higgs sector.

\section{ Higgs Search at the LHC }

The basic features that arise from the numerical calculations are that
the squark production cross section peak near the real production thresholds,
$\sqrt{s}\sim 2m_{\tilde{q}}$. This enhancement should be smoothed somewhat 
after the inclusion of QCD corrections.
The overall level of the squark loop contribution is much smaller than that 
of the quark loops, and even after including the interference between the 
two contributions we find that the squark loops can always be safely neglected
in an overall estimate of the background level. 

In Fig.~7, we present the invariant mass distribution of $ZZ$ 
including the Higgs signal from gluon fusion and the irreducible background 
from $q\overline{q}$ annihilation and $gg$ fusion via quark and squark loops 
in $pp$ collisions, with a rapidity cut $|y_Z| < 2.5$, 
for $m_{\tilde{q}} = 200$ GeV, $\tan\beta = 2$ and $\tan\beta = 10$.
For the smaller values of  $\tan\beta$ near 2, pronounced 
peaks appear at $M_{ZZ} = M_H$.
For $\tan\beta \agt 10$, the Higgs peaks become almost invisible.

The Higgs signal in the ``gold-plated'' mode 
$H\to ZZ \to 4\ell$ is viable provided there are no
supersymmetric decays of the Higgs boson \cite{Z2Z2}. 
The ``silver-plated'' mode $H\to ZZ \to \ell^+\ell^-\nu\overline{\nu}$ 
might also be used, and has a rate six times that of the gold-plated mode, 
but reducible $Z$-jet backgrounds might be a problem \cite{silver,LHC}.
If real squarks can be produced the Higgs width becomes larger, the Higgs peak 
in the invariant mass of $ZZ$ is reduced and the event rate 
is too small to discover the Higgs. 
Even if we choose squarks with a mass such that the squark contribution 
peaks near the Higgs pole, the contribution to the continuum background 
from the squark loops is negligible. 

\begin{center}
\epsfxsize=4.5in
\hspace*{0in}
\epsffile{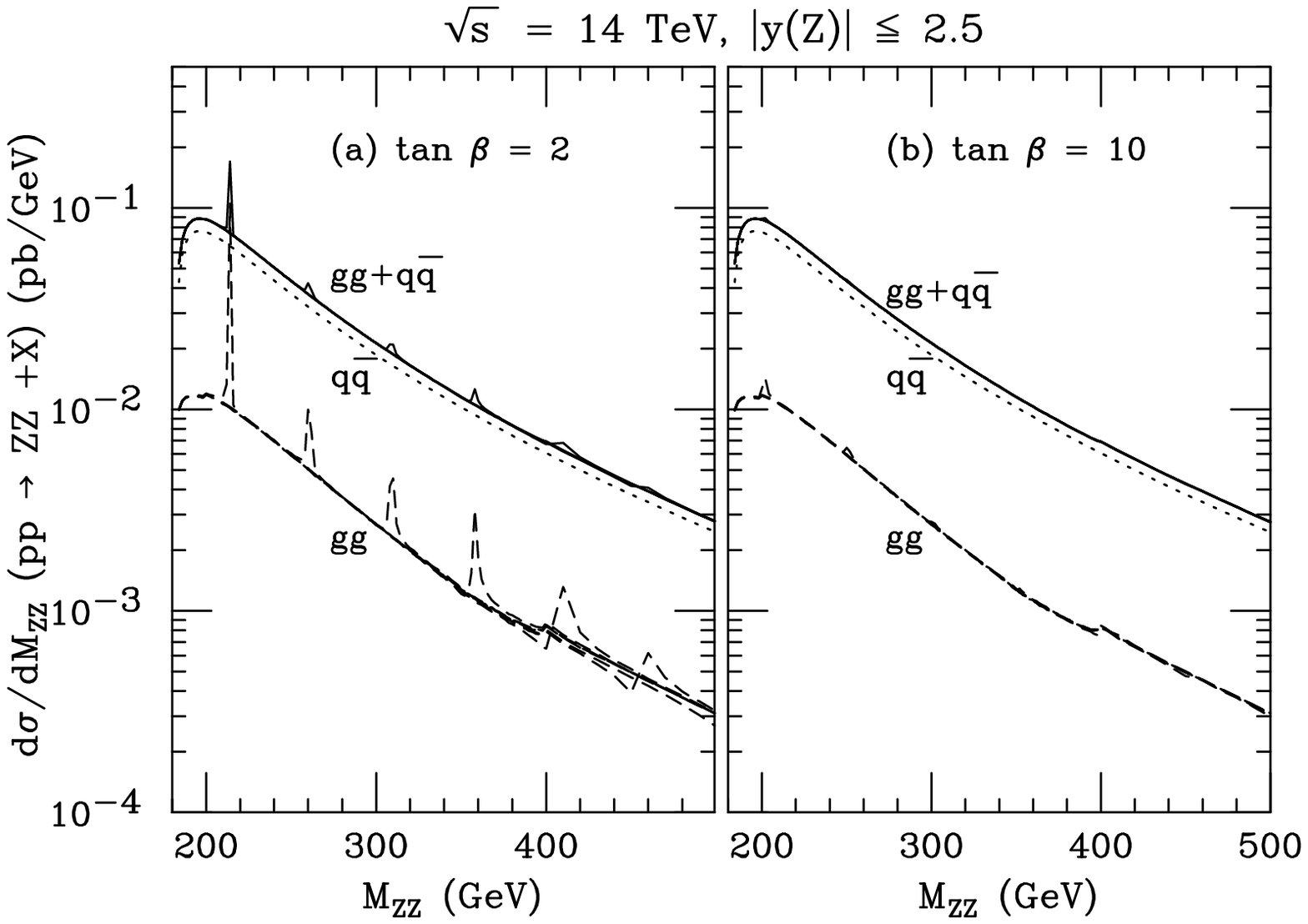}
\vspace*{-2.5in}

\smallskip
\parbox{5.5in}{\small  Fig.~7. Differential cross section of the Higgs signal 
in $pp$ collisions with a rapidity cut $|y_Z^{} < 2.5|$,
from quark and squark loops, for $m_{\tilde{q}} = 200$ GeV,
as a function of $M_{ZZ}$, for $\tan \beta = 2$ and $10$;
and, the irreducible background $d\sigma /dM_{ZZ}$ via 
$q\overline{q}$ and $gg$ fusion.}
\end{center}
\vspace*{0.2in}

We expect that the squark loop background is also negligible for Higgs bosons
with masses $M_{H,h}<2M_Z$, so that at least one of the $Z$ bosons is 
virtual $(Z^*)$. 
In fact for the lighter supersymmetric Higgs, $h$, this is an attractive 
channel for discovery at the LHC \cite{gsw}.
The calculation $gg\to ZZ^*\to 4\ell$ has been performed for the quark 
loops in Ref.~\cite{zmvdb}, and that background falls rapidly for invariant 
masses decreasing below the $ZZ$ threshold. We expect the squark loops to 
be negligible in this case as well, but our calculation is applicable only 
for real $Z$ bosons.

\section{ Conclusions }

We have calculated the contribution from squark loops to the process 
$gg\to ZZ$ at the LHC. We find that they are substantially smaller than the 
contribution from quark loops, and can usually be safely neglected in 
computation of the continuum background in the Higgs search.
In models with two CP-even Higgs bosons, we have also shown 
the amplitudes involving the heavy Higgs boson 
interferes destructively just below the resonance with the contributions
from the light Higgs bosons.
The expressions for the helicity amplitudes given in the appendix 
can be easily adapted to give the squark loop contributions to the processes 
$gg\to \gamma \gamma$, $\gamma \gamma \to \gamma \gamma$ and 
$\gamma \gamma \to ZZ$. 
The first process is relevant for Higgs searches at the LHC, 
while the last two are relevant for the $\gamma \gamma $ options 
of future linear $e^+e^-$ colliders.

\section*{ acknowledgements }

This research was supported in part by the U.S. Department of Energy 
under Grants No. DE-FG02-95ER40661 and No. DE-FG02-95ER40896
and in part by the University of Wisconsin Research Committee
with funds granted by the Wisconsin Alumni Research Foundation.

\section*{ appendix }

In this appendix we give the formulas for the helicity amplitudes.
We define helicity vectors for the process ($12\to 34$) as
\begin{eqnarray}
e_1^+&=&e_2^-={1\over \sqrt{2}}(0;-i,1,0)\;, \\
e_1^-&=&e_2^+={1\over \sqrt{2}}(0;i,1,0)\;, \\
e_3^+&=&e_4^-={1\over \sqrt{2}}(0;i\cos \theta,1,-i\sin \theta)\;, \\
e_3^-&=&e_4^+={1\over \sqrt{2}}(0;-i\cos \theta,1,i\sin \theta)\;, \\
e_3^0&=&{1\over m_Z}(q;p\sin \theta,0,p\cos \theta)\;, \\
e_4^0&=&{1\over m_Z}(q;-p\sin \theta,0,-p\cos \theta)\;, 
\end{eqnarray}
where the momenta are defined as
\begin{eqnarray}
p_1^\mu &=& (p;0,0,p)\;, \\
p_2^\mu &=& (p;0,0,-p)\;, \\
p_3^\mu &=& (p;q\sin \theta,0,q\cos \theta)\;, \\
p_4^\mu &=& (p;-q\sin \theta,0,-q\cos \theta)\;,
\end{eqnarray}
in the $gg$ rest frame. One then can define the Mandelstam variables
\begin{eqnarray}
&&s=(p_1+p_2)^2\;, \quad t=(p_1-p_3)^2\;, \quad u=(p_1-p_4)^2\;.
\end{eqnarray}

In nonsupersymmetric extensions of the Higgs sector the adjustment to the 
calculations is clearly only in the diagrams containing the Higgs bosons for
which the single s-channel contribution is replaced by multiple s-channel 
Higgs contributions with couplings that preserve the unitarity cancellations 
that occur in the Standard Model. Indeed in the MSSM the unitarity cancellation
in the quark loop diagrams alone manifests itself as the contribution for the
$h$ and $H$ Higgs bosons. The Standard Model contribution from quark loops 
to the production 
of longitudinal pairs (apart from an overall factor) 
\begin{eqnarray}
{\cal A}_{++00}^{\rm Higgs}&=&{{2m_q^2}\over {s-m_H^2+i\Gamma _Hm_H}}
{{s_2}\over {2\mz ^2}}\left (-2+s_4C_0(s)\right )\;, 
\end{eqnarray}
becomes
\begin{eqnarray}
{\cal A}_{++00}^{\rm Higgs}&=&
{{s_2}\over {2\mz ^2}}\left (-2+s_4C_0(s)\right )\\ \nonumber
&&\times \left [{{2m_q^2}\over {s-m_h^2+i\Gamma _hm_h}}
{{\cos \alpha \sin (\beta -\alpha)}\over {\sin \beta}}
+{{2m_q^2}\over {s-m_H^2+i\Gamma _Hm_H}}
{{\sin \alpha \cos (\beta -\alpha)}\over {\sin \beta}}\right ]\;, 
\end{eqnarray} 
for up-type quarks, and 
\begin{eqnarray}
{\cal A}_{++00}^{\rm Higgs}&=&
{{s_2}\over {2\mz ^2}}\left (-2+s_4C_0(s)\right )\\ \nonumber
&&\times \left [-{{2m_q^2}\over {s-m_h^2+i\Gamma _hm_h}}
{{\sin \alpha \sin (\beta -\alpha)}\over {\cos \beta}}
+{{2m_q^2}\over {s-m_H^2+i\Gamma _Hm_H}}
{{\cos \alpha \cos (\beta -\alpha)}\over {\cos \beta}}\right ]\;, 
\end{eqnarray} 
for down-type quarks
in the MSSM. 

Extracting an overall factor 
\begin{eqnarray}
{{2\alpha_s\alpha}\over {\sin ^2\theta _w\cos ^2\theta _w}} \left (L^2
+R^2\right )\label{factor}
\end{eqnarray}
where
\begin{eqnarray}
L&=&{1\over 2}-{2\over 3}\sin ^2\theta_w\;, \\
R&=&{2\over 3}\sin ^2\theta_w\;, 
\end{eqnarray}
for charge $2/3$ squarks and where
\begin{eqnarray}
L&=&-{1\over 2}+{1\over 3}\sin ^2\theta_w\;, \\
R&=&-{1\over 3}\sin ^2\theta_w\;, 
\end{eqnarray}
for charge $-1/3$ squarks,
the helicity amplitudes for squark loop diagrams for the process
$gg\rightarrow ZZ$ not involving the Higgs 
bosons are:

\begin{eqnarray}
{\cal A}_{+-++}&=&\Bigg \{D(s,t)\left ({{st^2}\over {Y}}+2\ms^2\right )
\left (\ms^2+{{\mz^4}\over {s_4}}\right )\nonumber \\
&&+C(t)\left [{{2\ms^2s_2}\over {s_4}}\left ({{u_1}\over s}
-{\mz^2\over {t_1}}\right )-2{{tt_1}\over {Y}}\left (\ms^2+
{{\mz^4}\over {s_4}}\right )\right ]
\nonumber \\
&&+B(t){{\mz^2}\over {s_4t_1^2}}(Y+2\mz^2t)+(t\leftrightarrow u)\Bigg \}
\nonumber \\
&&+D(t,u)\ms ^2\left (2\ms^2+{{s_2Y}\over {ss_4}}\right )\nonumber \\
&&+C_0(s)\left ({{ss_2}\over {Y}}\right )\left (\ms^2+{{\mz ^4}\over {s_4}}
\right )\nonumber \\
&&+C(s)\left ({{t^2+u^2-2\mz^4}\over {Y}}\right )
\left (\ms^2+{{\mz ^4}\over {s_4}}
\right )\nonumber \\
&&-{{Y}\over {t_1u_1}}\left (1+2{{\mz^2}\over {s_4}}\right )\;,
\end{eqnarray}

\begin{eqnarray}
{\cal A}_{+++-}&=&\Bigg \{D(s,t)\left [2\ms^2\left (\ms^2-{{\mz^4}\over {s_4}}
\right )+{{\ms^2st^2}\over {Y}}\right ]\nonumber \\
&&+C(t)\left [2{{\mz^2t_1Y}\over {s^2s_4}}-2\ms^2\left [1+
{{\mz^2s_2}\over {s_4t_1}}+{t_1 \over s}\right ]-2\ms^2{{tt_1}\over Y}\right ]
\nonumber \\
&&-B(t)\mz^2\left(2{t\over s}-1\right ){{Y}\over {s_4t_1^2}}
+(t\leftrightarrow u)\Bigg \}
\nonumber \\
&&+D(t,u)\left ({Y\over s}+2\ms^2\right )
\left (\ms^2-{{Y\mz^2}\over {ss_4}}\right )\nonumber \\
&&+C_0(s)\left ({{s_2}\ms^2\over {s_4Y}}\right )\left (t^2+u^2-2\mz^4
\right )\nonumber \\
&&+C(s){{ss_4\ms^2}\over {Y}}\nonumber \\
&&-{{Y}\over {t_1u_1}}\left (1+2{{\mz^2}\over {s_4}}\right )\;,
\end{eqnarray}

\begin{eqnarray}
{\cal A}_{+-00}&=&\Bigg \{D(s,t){{s\mz^2}\over {s_4}}\left (
\ms^2+{{st^2}\over {2Y}}\right )\nonumber \\
&&+C(t){{\mz^2}\over {s_4}}\left [8{{\ms^2Y}\over {st_1}}-{{stt_1}\over {Y}}
\right ] 
\nonumber \\
&&-B(t)2{{m_Z^2}\over {s_4t_1^2}}(t^2+\mz^4)
+(t\leftrightarrow u)\Bigg \}
\nonumber \\
&&-D(t,u){{(t-u)^2\mz^2\ms^2}\over {ss_4}}\nonumber \\
&&+C_0(s)\left ({{s^2s_2\mz^2}\over {2s_4Y}}\right )\nonumber \\
&&+C(s)\left ({{s\mz^2}\over {2s_4Y}}\right )\left (t^2+u^2-2\mz^4\right )
\nonumber \\
&&-{{4\mz^2Y}\over {s_4t_1u_1}}\;,
\end{eqnarray}

\begin{eqnarray}
{\cal A}_{++00}&=&\Bigg \{D(s,t){{s\mz^2\ms^2}\over {s_4}}
\nonumber \\
&&+C(t)\mz^2\left ({{(t-u)^2t_1}\over {s^2s_4}}+8{{\mz^2\ms^2}\over {s_4t_1}}
\right )
\nonumber \\
&&-B(t){{\mz^2}\over {s_4}}\left (4{t\over s}+2-4{{\mz^4}\over {t_1^2}}\right )
+(t\leftrightarrow u)\Bigg \}
\nonumber \\
&&-D(t,u){{\mz^2(t-u)^2}\over {2s^2s_4}}\left (Y+2\ms^2s\right )\nonumber \\
&&+C_0(s)\left (8{{\mz^2\ms^2}\over {s_4}}\right )\nonumber \\
&&+{{4\mz^2Y}\over {s_4t_1u_1}}\;,
\end{eqnarray}

\begin{eqnarray}
{\cal A}_{+-+-}&=&\Bigg \{D(s,t)\Bigg [{1\over {2s_4Y^2}}\left (
2\ms^2Y+st^2\right )
\left (2\ms^2s_4Y+ss_4t^2-2\mz^4Y\right )\nonumber \\
&&+\beta_s {{st}\over {2s_4Y^2}}\left (4\ms^2Y+st^2\right )
\left (t^2-\mz^4+Y\right )\Bigg ]
\nonumber \\
&&+C(t)\Bigg [2\ms^2{1\over {ss_4t_1Y}}\left (
\mz^2Y(t^2+u^2-2\mz^4)-ss_4tt_1^2\right )+{{tt_1}\over {s_4Y^2}}
\left (2\mz^4Y-ss_4t^2\right )\nonumber \\
&&+\beta_s\left \{2\ms^2{{s}\over {s_4t_1Y}}\left (t(t^2-\mz^4)-\mz^2Y\right )
-{{st^2t_1}\over {s_4Y^2}}(t^2-\mz^4+Y)\right \}\Bigg ]
\nonumber \\
&&+B(t)\Bigg [2{{\mz^4}\over {s_4t_1}}+{{\mz^4}\over {t_1^2}}\left (
1+2{{\mz^2}\over {s_4}}\right )-{1\over 2}+{{\mz^2}\over {s_4}}\nonumber \\
&&-\beta_s\left (2{{\mz^4}\over {s_4t_1}}+\mz^4{{s}\over {s_4t_1^2}}
+2{t\over {s_4}}-{1\over 2}\right )-{{t^2}\over Y}\left (
1+{{\beta_s (t-u)}\over {s_4}}\right )\Bigg ]
+(t\leftrightarrow u,\beta_s \rightarrow -\beta_s)\Bigg \}
\nonumber \\
&&+D(t,u)2\ms^2\left (\ms^2-{{\mz^2Y}\over {ss_4}}\right )\nonumber \\
&&+C_0(s){{ss_2}\over Y}\Bigg [\ms^2+{{[(t^2+u^2-2\mz^4)(t^2+tu+u^2-2\mz^4)
-2Y^2]}\over {2s_4Y}}\nonumber \\
&&+\beta_s {{t-u}\over {s_4}}\Bigg \{\ms^2
+{{s(t^2+tu+u^2-2\mz^4)}\over {2Y}}\Bigg \}\Bigg ]
\nonumber \\
&&+C(s){s\over Y}\Bigg [\ms^2s_4+{{t^2(t^2-\mz^4+Y)
+u^2(u^2-\mz^4+Y)}\over {2Y}}
\nonumber \\
&&+\beta_s (t-u)\Bigg \{\ms^2+{{t^2(t^2-\mz^4+Y)+u^2(u^2-\mz^4+Y)
+2(s_2^2-\mz^4)Y}
\over {2s_4Y}}\Bigg \}\Bigg ]\nonumber \\
&&+B(s)\left [\left ({{s_2(t^2+u^2-2\mz^4+Y)}\over {s_4Y}}\right )
+\beta_s\left ({{ss_2(t-u)}\over {s_4Y}}\right )\right ]\nonumber \\
&&+{1\over {s_4t_1u_1}}\left [s_2Y+\beta_s \mz^2s(t-u)\right ]\;,
\end{eqnarray}

\begin{eqnarray}
{\cal A}_{+++0}/\Delta&=&\Bigg \{D(s,t){{s\ms^2}\over {2s_4}}\left (t^2-\mz^4+Y
\right )
\nonumber \\
&&+C(t)\Bigg [\ms^2{{s(t^2-\mz^4+Y)}\over {s_4t_1}}+{{(t-u)t_1Y}\over {2ss_4}}
\nonumber \\
&&+\beta_s\left \{\ms^2{{s(t^2-\mz^4+Y)+t_1^2(t-u)}\over {s_4t_1}}
+{{t_1(t-u)Y}\over {2ss_4}}
\right \}\Bigg ]
\nonumber \\
&&+B(t)(1+\beta_s){{sY}\over {2s_4}}\left [{{\mz^2}\over {t_1}}\left (
{1\over {t_1}}-{2\over s}\right )-{2\over s}\right ]
-(t\leftrightarrow u)\Bigg \}
\nonumber \\
&&-D(t,u){{Y(t-u)}\over {4ss_4}}\left [\left (Y+2s\ms^2\right )+\beta_s 
\left (Y+4s\ms^2\right )\right ]\nonumber \\
&&-C_0(s){{s\ms^2(t-u)}\over {s_4}}\nonumber \\
&&-(1+\beta_s){{s(t-u)Y}\over {2s_4t_1u_1}}\;,
\end{eqnarray}

\begin{eqnarray}
{\cal A}_{+-+0}/\Delta&=&\Bigg \{D(s,t)\Bigg [
{s\over {4s_4Y}}\left (t^2-\mz^4+Y
\right )\left (2\ms^2Y+st^2\right )\nonumber \\
&&+\beta_s{{s^2t}\over {4Ys_4}}\left (4\ms^2Y+st^2\right )\Bigg ]
\nonumber \\
&&-C(t)\Bigg [\ms^2{{(t-u)Y}\over {s_4t_1}}+{{stt_1(t^2-\mz^4+Y)}\over {2s_4Y}}
\nonumber \\
&&-\beta_s \left \{\ms^2{{s(st+2Y)}\over {s_4t_1}}-{{s^2t^2t_1}\over {2s_4Y}}
\right \}\Bigg ]
\nonumber \\
&&+B(t){s\over {2s_4}}\left [\mz^2(t-u){t\over {t_1^2}}+t+\mz^2-\beta_s\left (
\mz^4{s\over {t_1^2}}+\mz^2{s\over t_1}+t+\mz^2\right )\right ]
\nonumber \\
&&-(t\leftrightarrow u, \beta_s \rightarrow -\beta_s)\Bigg \}
\nonumber \\
&&-D(t,u){{\ms^2Y(t-u)}\over {2s_4}}\nonumber \\
&&-C_0(s){{s^2}\over {4s_4}}\Bigg [{{(t-u)(t^2+u^2+2Y)}\over Y}
+\beta_s \left (4\ms^2+{{s(t^2+u^2)}\over Y}\right )\Bigg ]\nonumber \\
&&-C(s){{s^2s_2}\over {4Y}}\Bigg [(t-u)+\beta_s{{t^2+u^2-2\mz^4}\over {s_4}}
\Bigg ]
\nonumber \\
&&-B(s){{s(t-u+\beta_s s)}\over {2s_4}}\nonumber \\
&&+{{sY(t-u-s\beta_s)}\over {2s_4t_1u_1}}\;,
\end{eqnarray}

\begin{eqnarray}
{\cal A}_{++++}&=&\Bigg \{D(s,t)2\ms^2\left (\ms^2+{{\mz^4}\over {s_4}}
\right )
\nonumber \\
&&+C(t)\Bigg [-2{{\ms^2}\over {ss_4t_1}}\left (
s_2(t_1(t-u)+Y)-2\mz^2t_1^2\right )-{{s_2t_1Y}\over {s^2s_4}}\nonumber \\
&&+\beta_s \left \{-2\ms^2{{t_1(t-u)+Y}\over {s_4t_1}}-{{t_1Y}\over {ss_4}}
\right \}\Bigg ]
\nonumber \\
&&+B(t)\left ({{(1+\beta_s)s-2\mz^2}\over {2s_4}}\right )
\left (2{{\mz^4}\over {t_1^2}}-2{t\over s}-1\right )
+(t\leftrightarrow u)\Bigg \}
\nonumber \\
&&+D(t,u){1\over {2s^2s_4}}\left [\left (2s\ms^2+Y\right )
\left (2ss_4\ms^2+s_2Y\right )+\beta_s sY\left (Y+4s\ms ^2\right )\right ]
\nonumber \\
&&+C_0(s)4{{\mz^2\ms^2}\over {s_4}}\nonumber \\
&&+{1\over {s_4t_1u_1}}\left [\mz^2(2Y-ss_4)+\beta_s sY\right ]\;,
\end{eqnarray}
where $s_4=s-4\mz^2$, $s_2=s-2\mz^2$, $t_1=t-\mz^2$, $u_1=u-\mz^2$, 
$\beta_s=\sqrt{s_4/s}$, $Y=tu-\mz^4$, and $\Delta=\sqrt{-2\mz^2/sY}$.
The squark mass-squared is denoted by $\ms$, where this is the 
eigenvalue(s) 
of the squark mass-squared matrix. Different masses for the left- and 
right-handed squarks can be accounted for by separating the overall factors in 
Eq.~\ref{factor},and using two squark masses $m_{\tilde{q}_L}^2$ and 
$m_{\tilde{q}_R}^2$
in the expressions for the helicity amplitudes above.
The factors $B$, $C$ and $D$ are the usual scalar integrals 
that result from the reduction \cite{Tini,FF} of the tensor integrals in the 
original Feynman diagrams. Expressions for these in terms of logarithms and 
dilogarithms can be found in Ref.~\cite{ggzz2}.
The other helicity amplitudes are related by the same substitutions required
for the quark loops \cite{ggzz2}
\begin{eqnarray}
{\cal A}_{\lambda_1\lambda_2\lambda_3\lambda_4}&=&
{\cal A}^*_{-\lambda_1-\lambda_2-\lambda_3-\lambda_4}\;, \nonumber \\
{\cal A}_{++--}(\beta)&=&{\cal A}_{++++}(-\beta)\;, \nonumber \\ 
{\cal A}_{+++-}(\beta)&=&{\cal A}_{++-+}(\beta)\;, \nonumber \\
{\cal A}_{+---}(\beta)&=&{\cal A}_{+-++}(\beta)\;, \nonumber \\
{\cal A}_{+--+}(\beta)&=&{\cal A}_{+-+-}(-\beta)\;, \nonumber \\
{\cal A}_{+++0}(\beta)={\cal A}_{++0+}(\beta)&=&
{\cal A}_{++-0}(-\beta)={\cal A}_{++0-}(-\beta)\;, \nonumber \\
{\cal A}_{+-+0}(\beta)=-{\cal A}_{+-0+}(-\beta)&=&
{\cal A}_{+--0}(-\beta)=-{\cal A}_{+-0-}(\beta)\;.
\end{eqnarray}

The helicity amplitudes for top 
squark loop diagrams containing the Higgs bosons 
for the process
$gg\rightarrow ZZ$ are:

\begin{eqnarray}
{\cal A}_{++00}^{\rm Higgs}&=&{{\alpha_s\alpha}\over 
{\sin ^2\theta _w\cos ^2\theta _w}}\Bigg [\left ({{2m_t^2}\over {\mw}}
{{\sin \alpha}\over {\sin \beta}}+{{\mz}\over {2\cos \theta _w}}
\cos(\beta +\alpha )\right ){{\mz }\over {\cos \theta _w}}\nonumber \\
&&\quad \quad \times {s_2\over {2\mz ^2}}(1+2\ms ^2C_0(s))
{{\cos (\beta -\alpha )}\over {s-m_H^2+i\Gamma_Hm_H}}\nonumber \\
&&+\left ({{2m_t^2}\over {\mw}}
{{\cos \alpha}\over {\sin \beta}}-{{\mz}\over {2\cos \theta _w}}
\sin(\beta +\alpha )\right ){{\mz }\over {\cos \theta _w}}\nonumber \\
&&\quad \quad \times {s_2\over {2\mz ^2}}(1+2\ms ^2C_0(s))
{{\sin (\beta -\alpha )}\over {s-m_h^2+i\Gamma_hm_h}}\Bigg ]\;, \\
{\cal A}_{++++}^{\rm Higgs}&=&{{\alpha_s\alpha}\over 
{\sin ^2\theta _w\cos ^2\theta _w}}\Bigg [\left ({{2m_t^2}\over {\mw}}
{{\sin \alpha}\over {\sin \beta}}+{{\mz}\over {2\cos \theta _w}}
\cos(\beta +\alpha )\right ){{\mz }\over {\cos \theta _w}}\nonumber \\
&&\quad \quad \times (1+2\ms ^2C_0(s))
{{\cos (\beta -\alpha )}\over {s-m_H^2+i\Gamma_Hm_H}}\nonumber \\
&&+\left ({{2m_t^2}\over {\mw}}
{{\cos \alpha}\over {\sin \beta}}-{{\mz}\over {2\cos \theta _w}}
\sin(\beta +\alpha )\right ){{\mz }\over {\cos \theta _w}}\nonumber \\
&&\quad \quad \times (1+2\ms ^2C_0(s))
{{\sin (\beta -\alpha )}\over {s-m_h^2+i\Gamma_hm_h}}\Bigg ]\;, 
\end{eqnarray}
and the contributions from the bottom squark are:
\begin{eqnarray}
{\cal A}_{++00}^{\rm Higgs}&=&{{\alpha_s\alpha}\over 
{\sin ^2\theta _w\cos ^2\theta _w}}\Bigg [\left ({{2m_b^2}\over {\mw}}
{{\cos \alpha}\over {\cos \beta}}-{{\mz}\over {2\cos \theta _w}}
\cos(\beta +\alpha )\right ){{\mz }\over {\cos \theta _w}}\nonumber \\
&&\quad \quad \times {s_2\over {2\mz ^2}}(1+2\ms ^2C_0(s))
{{\cos (\beta -\alpha )}\over {s-m_H^2+i\Gamma_Hm_H}}\nonumber \\
&&+\left (-{{2m_b^2}\over {\mw}}
{{\sin \alpha}\over {\cos \beta}}+{{\mz}\over {2\cos \theta _w}}
\sin(\beta +\alpha )\right ){{\mz }\over {\cos \theta _w}}\nonumber \\
&&\quad \quad \times {s_2\over {2\mz ^2}}(1+2\ms ^2C_0(s))
{{\sin (\beta -\alpha )}\over {s-m_h^2+i\Gamma_hm_h}}\Bigg ]\;, \\
{\cal A}_{++++}^{\rm Higgs}&=&{{\alpha_s\alpha}\over 
{\sin ^2\theta _w\cos ^2\theta _w}}\Bigg [\left ({{2m_b^2}\over {\mw}}
{{\cos \alpha}\over {\cos \beta}}-{{\mz}\over {2\cos \theta _w}}
\cos(\beta +\alpha )\right ){{\mz }\over {\cos \theta _w}}\nonumber \\
&&\quad \quad \times (1+2\ms ^2C_0(s))
{{\cos (\beta -\alpha )}\over {s-m_H^2+i\Gamma_Hm_H}}\nonumber \\
&&+\left (-{{2m_b^2}\over {\mw}}
{{\sin \alpha}\over {\cos \beta}}+{{\mz}\over {2\cos \theta _w}}
\sin(\beta +\alpha )\right ){{\mz }\over {\cos \theta _w}}\nonumber \\
&&\quad \quad \times (1+2\ms ^2C_0(s))
{{\sin (\beta -\alpha )}\over {s-m_h^2+i\Gamma_hm_h}}\Bigg ]\;. 
\end{eqnarray}

We remark here that the equations above involving transverse $Z$ bosons 
can be reduced to the helicity 
amplitudes for $gg\to \gamma \gamma$ from squark loops not involving the 
Higgs graphs (the process $gg\to h\to \gamma \gamma $ has been investigated
including squark mixing in Refs.~\cite{Kileng,Djouadi}). 
Also the contribution to 
photon-photon scattering $\gamma \gamma \to \gamma \gamma$ can be obtained 
from that for $gg\to \gamma \gamma$ by changing overall factors.



\begin{references}

\bibitem{bbps} V.~Barger, M.S.~Berger, R.J.N.~Phillips and A.~Stange,
Phys.\ Rev.\ {\bf D45}, 4128 (1992); 
J.F.~Gunion, R.~Bork, H.E.~Haber and A.~Seiden, Phys.\ Rev.\ {\bf D46}, 
2040 (1992);
Z.~Kunszt and F.~Zwirner, Nucl.\ Phys.\ {\bf B385}, 3 (1992).

\bibitem{Z2Z2}
H.~Baer, M.~Bisset, D.~Dicus, C.~Kao and X.~Tata, 
Phys. Rev. {\bf D47} (1993) 1062; 
H.~Baer, M.~Bisset, C.~Kao and X.~Tata, Phys. Rev. {\bf D50} (1994) 316.

\bibitem{LHC}
CMS Technical Proposal, CERN/LHCC 94-38 (1994); 
Atlas Technical Proposal, CERN/LHCC 94-43 (1994); 
E.~Richter-Was, et al., Int. J. Mod. Phys. {\bf A13}, 1371 (1998). 

\bibitem{gsw} A summary of the many discovery possibilities for weakly
coupled Higgs bosons can be found in J.~F.~Gunion, A.~Stange and 
S.~Willenbrock, UCD-95-28, hep-ph/9602238. 

\bibitem{Kileng} B.~Kileng, Z.\ Phys.\ {\bf C63}, 87 (1994).

\bibitem{Spira} M.~Spira, A.~Djouadi, D.~Graudenz and P.M.~Zerwas,
Nucl.\ Phys.\ {\bf B453}, 17 (1995); S.~Dawson, A.~Djouadi, and M.~Spira,
Phys.\ Rev.\ Lett.\ {\bf 77}, 16 (1996). 

\bibitem{ggzz1} D.A.~Dicus, C.~Kao,
and W.W.~Repko, Phys.\ Rev.\ {\bf D36} 1570 (1987).

\bibitem{ggzz2} E.W.N.~Glover and J.J.~van~der~Bij, Phys.\ Lett.\
{\bf B219}, 488 (1989); Nucl.\ Phys.\ {\bf B321}, 561 (1989).

\bibitem{Konig} H.~K\"{o}nig, Z.\ Phys.\ {\bf C69}, 493 (1996).

\bibitem{Djouadi} A.~Djouadi, hep-ph/9806315.

\bibitem{Tini} 
G.~`t~Hooft and M.~Veltman, Nucl. Phys. {\bf B153} (1979) 365;
G.~Passarino and M.~Veltman, Nucl. Phys. {\bf B160} (1979) 151. 

\bibitem{FF} G.J.~van~Oldenborgh and J.A.M.~Vermaseren, 
Z.\ Phys.\ {\bf C46}, 425 (1990).

\bibitem{LOOP} 
D.A. Dicus and C. Kao, LOOP, 
a FORTRAN program for evaluating one-loop integrals. 

\bibitem{prep} M.S.~Berger and C.~Kao, in preparation.

\bibitem{CTEQ}  H.-L.~Lai, et al., Phys.\ Rev.\  {\bf D51}, 4763 (1995).

\bibitem{squark} CDF Collaboration, F.~Abe et al.,
Phys.\ Rev.\ {\bf D56}, 1357 (1997);
D0 Collaboration, A.~Abachi et al., 
Phys.\ Rev.\ Lett.\ {\bf 75}, 618 (1995).

\bibitem{susy1} Z.~Bern and D.A.~Kosower, Nucl.\ Phys.\ {\bf B379}, 451 
(1992); ibid. {\bf B379}, 562 (1992); 
C.S.~Lam, Nucl.\ Phys.\ {\bf B397}, 143 (1993); 
Z.~Bern, L.~Dixon and D.A.~Kosower, Phys.\ Rev.\ Lett.\ {\bf 70}, 2677 (1993).

\bibitem{susy2} Z.~Bern and A.G.~Morgan, Phys.\ Rev.\ {\bf D49},  6155 (1994).

\bibitem{Jikia} G.V.~Jikia and A.~Tkabladze, 
Phys.\ Lett.\ {\bf B323}, 453 (1994).

\bibitem{ppHH} E.W.N.~Glover and J.J.~van~der~Bij, 
Nucl.\ Phys.\ {\bf B309}, 282 (1988).

\bibitem{HH} G.V.~Jikia, Nucl.\ Phys.\ {\bf B412}, 57  (1994).

\bibitem{ppzz} G.V.~Jikia, Phys.\ Lett.\ {\bf B298}, 224 (1993); Nucl.\
Phys.\ {\bf B405}, 24  (1993); 
M.S.~Berger, Phys.\ Rev.\ {\bf D48}, 5121 (1993); 
D.A.~Dicus and C.~Kao, Phys.\ Rev.\ {\bf D49}, 1265 (1994).

\bibitem{silver} M.S.~Chanowitz and M.K.~Gaillard, 
Nucl.\ Phys.\ {\bf B261}, 379 (1985); M.S.~Chanowitz and R.N.~Cahn,
Phys.\ Rev.\ Lett.\ {\bf 56}, 1327 (1986).

\bibitem{zmvdb} C.~Zecher, T.~Matsuura and J.J.~van~der~Bij, 
Z.\ Phys.\ {\bf C64}, 219 (1994). 

\end{references}
\end{document}